\documentclass[letterpaper]{article} 
\usepackage{aaai24}  
\usepackage{times}  
\usepackage{helvet}  
\usepackage{courier}  
\usepackage[hyphens]{url}  
\usepackage{graphicx} 
\usepackage{natbib}  
\usepackage{caption} 
\usepackage{algorithm}
\usepackage{amsmath}
\usepackage{amsfonts}
\usepackage{amssymb}
\usepackage{amsthm}
\usepackage{subfigure}
\usepackage{algpseudocode}
\newtheorem{proposition}{Proposition}

\usepackage{hyperref}

\title{Harnessing Network Effect for Fake News Mitigation: Selecting Debunkers via Self-Imitation Learning}

\author{
    Xiaofei Xu, Ke Deng, Michael Dann, Xiuzhen Zhang
}
\affiliations{
    School of Computing Technologies, RMIT University, Melbourne, Victoria 3000 Australia\\
    xiaofei.xu2@student.rmit.edu.au, \{ke.deng, michael.dann, xiuzhen.zhang\}@rmit.edu.au
}

\begin{document}
\maketitle
\begin{abstract}
This study aims to minimize the influence of fake news on social networks by deploying \textit{debunkers} to propagate true news. This is framed as a reinforcement learning problem, where, at each stage, one user is selected to propagate true news. A challenging issue is \textit{episodic reward} where the ``net'' effect of selecting individual debunkers cannot be discerned from the interleaving information propagation on social networks, and only the collective effect from mitigation efforts can be observed. Existing Self-Imitation Learning (SIL) methods have shown promise in learning from episodic rewards, but are ill-suited to the real-world application of fake news mitigation because of their poor sample efficiency. To learn a more effective debunker selection policy for fake news mitigation, this study proposes NAGASIL -- {\underline N}egative sampling and state {\underline A}ugmented {\underline Generative} {\underline Adversarial} {\underline S}elf-{\underline I}mitation {\underline L}earning, which consists of two improvements geared towards fake news mitigation: learning from negative samples, and an augmented state representation to capture the ``real" environment state by integrating the current observed state with the previous state-action pairs from the same campaign. Experiments on two social networks show that NAGASIL yields superior performance to standard GASIL and state-of-the-art fake news mitigation models. 
\end{abstract}
\section{Introduction}\label{sec:introduction}
There have been significant efforts to combat the spread of fake news on social networks. 
Beyond fake news detection, another important strategy is mitigation, whereby debunkers -- users who propagate true news -- are deployed to counter the spread of fake news. Unfortunately, despite the efforts of ``official" debunkers, such as fact-checking services and authoritative organisations (e.g., WHO), fake news still proliferates widely on social media. It is therefore important to unleash the power of crowd debunking from online users for fake news mitigation~\citep{vo2018rise,vo2020facts}. Previous work on this problem has explored using reinforcement learning to optimize fake news mitigation campaigns. Some studies focus on optimizing the intensity with which given debunkers spread true news~\citep{farajtabar2017fake,goindani2020cluster,goindani2020social}. Other studies optimize the selection of crowd debunkers from online users over multiple stages with a budget constraint~\citep{xiaofei2022}. 

A crucial issue in reinforcement learning-based multi-stage fake news mitigation strategies but one that is largely overlooked in the literature~\citep{farajtabar2017fake,goindani2020cluster,goindani2020social,xiaofei2022} is the lack of direct, ``net'' rewards for agent actions. Since news propagation on social networks is a long-lasting process with intensity decay over time, the effects of debunkers on fake news mitigation are interleaved across the network at different stages. The ``net" effect of individual debunkers cannot be directly measured at intermediate stages; only the cumulative effect of all debunkers can be observed when the mitigation campaign finishes. In other words, the reward function for the task is episodic.

In this paper, we propose a reinforcement learning approach to multi-stage fake news mitigation that explicitly addresses the issue of episodic reward. In a multi-stage fake news mitigation campaign, one user is selected at each stage to post true news such that the number of users believing in fake news is minimized at the conclusion of the campaign. To address the issue of episodic rewards, we propose to learn the debunker selection policy via self-imitation learning~\citep{oh2018self,gangwani2018learning}. Instead of learning directly from immediate rewards, self-imitation learning aims to mimic the agent's own past behaviour from highly rewarding episodes. Existing self-imitation learning algorithms ~\citep{ho2016generative,gangwani2018learning,oh2018self} have shown promise in learning from episodic rewards~\citep{gangwani2018learning}, but have two weaknesses that potentially limit their effectiveness for fake news mitigation on social networks:

First, fake news mitigation campaigns are cost-intensive \cite{farajtabar2017fake}. Real-world campaigns often have budget constraints and it is essential to learn mitigation policies from a small number of episodes. It is therefore desirable to exploit all available information about episodes for more efficient sampling and hence more efficient learning. Existing self-imitation learning methods imitate past \textit{good} experiences only, and state-action pairs that appear frequently in past good experiences are typically treated as favourable, but this may not be true if they also appear frequently in past bad experiences. Second, existing methods assume that the full environment state can be observed. However, in fake news mitigation, due to interleaved information propagation and the complexity of social networks, it is hard to observe users' reactions to news in real time.

In this paper, we argue that leveraging negative samples -- past experiences where the agent received a small episodic reward -- can yield more efficient sampling of good experiences and hence boost policy learning efficiency. Negative sampling informs the agent about undesirable behaviours and can therefore improve the efficiency of sampling good experiences. In addition, we devise an augmented state representation that better captures the ``true" environment state by integrating the current observations with previous state-action pairs in the same campaign. 

The contributions of this study are twofold: First, we propose, for the first time to our best knowledge, a multi-stage fake news mitigation approach designed for the realistic setting where rewards are episodic. Second, we propose NAGASIL -- Negative sampling and state Augmented Generative Adversarial Self-Imitation Learning -- which improves self-imitation learning for fake news mitigation via negative samples and augmented states. In addition to theoretically proving the advantage of negative samples and augmented states, we test NAGASIL on two social networks: one is a large synthetic network based on commonly used social network parameters, while the other is a widely used real-world social network for rumour propagation. Experiments demonstrate NAGASIL's superior performance compared to state-of-the-art fake news mitigation models\footnote{The source code for our experiments is available at https://github.com/xxfwin/NAGASIL}

\section{Related Work}\label{sec:literature}
There have been many studies on fake news detection on social networks. To reduce the cost and time burden of manual fact-checking, automated fact-checking of fake news and credibility analysis of social media posts have been proposed, using information such as network features \citep{benamira2019semi}, multi-modal features \citep{wang2018eann} and combined features \citep{shu2017fake,shu2019studying}. Other studies detect fake news and their spreaders on social networks based on linguistic and personality features~\citep{tian2020early,tian2022duck,shrestha2020detecting}. For a more complete review of automatic fact-checking systems, please refer to a recent survey~\cite{guo2022survey}.

Beyond fake news detection, research on strategies for propagating true news, such as fact-checked content, to mitigate the spread of fake news is attracting more attention. Various solutions have been investigated to select users as debunkers to propagate true news. These studies can be categorised along two lines: One line of research focuses on heuristics for one-off selection of debunkers for mitigation. These studies heuristically select the top-$k$ most influential users as debunkers \citep{saxena2020mitigating,saxena2020k}. They assume that users with high social influence will produce wide propagation of true news on social networks. However, research has shown that overall influence on social networks may not translate to wide mitigation propagation and reach users exposed to fake news as expected~\citep{farajtabar2016multistage}. 

Another line of research, which most strongly relates to ours, leverages reinforcement learning to optimize the cumulative effect across multiple stages of fake news mitigation~\citep{farajtabar2017fake,goindani2020cluster,goindani2020social,xiaofei2022}. In some studies~\citep{farajtabar2017fake,goindani2020cluster,goindani2020social}, a set of debunkers is given in advance, and each stage focuses on optimizing the intensity with which the debunkers post true news. Because the debunkers are fixed, even if they post with high intensity, the true news they propagate may not reach all users exposed to fake news, given the unknown origin and dynamic propagation of fake news on social networks \citep{xiaofei2022}. Not assuming fixed debunkers, \citeauthor{xiaofei2022} \citeyearpar{xiaofei2022} propose to select debunkers in a cost-effective way for multi-stage fake news mitigation. However, all of these studies assume that the mitigation effect of each stage can be immediately observed before the next stage, and overlook the issue of episodic reward that we described in the introduction, which can lead to less than optimal mitigation policies. In this paper, we address the critical issue of episodic reward, and our solution can be generally applied to other mitigation settings of selecting debunkers~\citep{xiaofei2022}. 

In addition to network-level mitigation, there are also studies on individual-level fake news mitigation~\citep{wang2022veracity,he2023reinforcement}.~\citeauthor{wang2022veracity} \citeyearpar{wang2022veracity} propose a personalised true news recommender system to counteract fake news by modelling both user interests and news veracity for individuals. With the recent advance of generative models~\citep{ouyang2022training}, a recent study focuses on generating personalized counter-misinformation posts~\citep{he2023reinforcement}. 

\section{Preliminaries and Problem Statement}\label{sec:3}
\textbf{Information Propagation Models.} Various information propagation models have been applied in research on fake news mitigation on social networks, including Hawkes processes \citep{farajtabar2017fake,lacombe2018fake,shu2019studying,goindani2020cluster,goindani2020social,murayama2021modeling}, Linear Threshold / Independent Cascade models \citep{pham2019minimum,saxena2020k}, Information Aggregation Games \citep{aymanns2019modeling}, and epidemic models \citep{zhao2013sir,wen2014shut,wen2014sword,tan2019aim}. Our proposed NAGASIL is independent of the propagation model. Without loss of generality, the epidemic model is employed as the environment for reinforcement learning in this study. 

To meet the specific requirements of different applications, several variants of the epidemic model have been proposed~\citep{brauer2008compartmental,gumel2004modelling}. In this study, we adopt SEIR (Susceptible-Exposed-Infected-Recovered). Let $X_{i}(t)$ be the \emph{epidemic state} (or \emph{e-state} for simplicity) of user $i$ at time $t$. $X_{i}(t)$ is always in one of four e-states: \textit{Susceptible}, \textit{Exposed}, \textit{Recovered}, or \textit{Infected}. $X_{i}(t)$ can transition from one e-state to another. If the user has not received any news yet, $X_{i}(t)$ is \textit{Susceptible}; if they have received fake/true news, $X_{i}(t)$ is \textit{Exposed}. Let $P_{i}^{I}$ ($P_{i}^{R}$) be the probability of user $i$ transitioning from some other e-state to \textit{Infected} (\textit{Recovered}).

\begin{equation}
P_{i}^{I}= 
\begin{cases}
\begin{aligned}
&L_i (N_i^F - N_i^M) &&\text{if}\ N_i^F>N_i^M \\ 
&0 &&\text{else}
\end{aligned}
\end{cases}\label{eq:infected}
\end{equation}
\begin{equation}
P_{i}^{R}=
\begin{cases}
\begin{aligned}
&L_i (N_i^M - N_i^F) && \text{if}\ N_i^M>N_i^F  \\ 
&0            && \text{else} 
\end{aligned}
\end{cases}\label{eq:recovered}
\end{equation}
where $N_i^F$ and $N_i^M$ are the number of fake and true news items received by user $i$ respectively, and $L_i(x)$ is defined as:
\begin{equation}\label{equ:lvalue}
L_{i}(x)=\frac{1}{1+e^{-\delta\left(x-x_{i}\right)}}
\end{equation}
where $x_i$ is the logistic function's midpoint, which is determined by the number of followers user $i$ has on the social network, $\delta$ is the logistic growth rate and $\delta=1$ by default. If a user has received more true news than fake news (i.e., $N_i^M>N_i^F$) and is not in e-state \textit{Recovered}, they will have a non-zero probability of transitioning to e-state \textit{Recovered}; otherwise, if they have received more fake news than true news (i.e., $N_i^M<N_i^F$) and are not in e-state \textit{Infected}, they will have a non-zero probability of transitioning to e-state \textit{Infected}. In both cases, the probability is lower if user $i$ has more followers. The rationale is that users with more followers tend to be harder to convince by news received~\citep{smit2022motivating}. 

Once a user transitions to e-state \textit{Infected} (or \textit{Recovered}), they have an initial intensity (i.e., probability) for spreading fake news (or true news). The intensity decays over time. Specifically, the intensity of user $i$ at time $t$ is: 
\begin{equation}\label{eq:intensity}
\iota_i(t) = \xi_i e^{-\omega (t-t_c)}.
\end{equation}
where $\xi_i$ is the initial intensity, $\omega$ controls the intensity decay rate ($\omega=1$ by default), and $t_c$ is the time when the e-state of user $i$ changed to \textit{Infected} (or \textit{Recovered}). The more time elapsed, the less likely the user is to spread fake/true news. 

Note that the users in the information propagation model continuously spread fake/true news based on their intensities. This leads to two consequences: First, the probability of each user transitioning to e-state \textit{Infected} (or \textit{Recovered}) changes over time, since it is determined by the number of fake/true news items received (Eq. \ref{eq:infected} and \ref{eq:recovered}). Second, the intensity of users spreading fake/true news changes continuously so it is difficult to observe intrinsic properties that drive users to interact with the environment (e.g., initial intensity).\\

\textbf{Problem Statement.} A social network is modelled as a directed graph $G(U, E)$ where $U$ and $E$ denote the social network users and the directed links between users, respectively. The information propagation on social networks is modelled by SEIR. Initially, the environment state is $s_0$, where the e-state of each user in $U$ is \textit{Susceptible}, \textit{Exposed} or \textit{Infected}, and the users in e-state \textit{Infected} have different intensities to spread fake news. $B$ is a budget and each user $i$ has a cost $c_i$. If user $i$ is selected as a debunker, the budget is reduced by $c_i$. Given a social network $G(U, E)$ with initial state $s_0$, the problem of selecting debunkers under budget $B$ for optimal mitigation can be mapped to a reinforcement learning problem, where the goal is to design a debunker selection policy such that, each $w$ time steps (i.e., a stage), the agent selects one user from $U$ as a new debunker based on the environment state at that time ($s_0$ for the first stage), and sets the user's e-state to \textit{Recovered} with the initial intensity to spread true news on $G$, until the remaining budget is less than the cost of any user in $U$. The optimization objective is to minimize the number of users in e-state \textit{Infected} in $U$ at time $t_f$ ($>\!\!>w$) after the multi-stage fake news mitigation campaign concludes. 

Since it is impractical to observe the full environment state in a realistic setting, we mask certain state features. The rationale is that we cannot observe the intrinsic properties that drive users to interact with the environment. The full environment state $s_E$ and the observed environment state $s$ are defined as follows:
\begin{equation}\label{eq5}
\begin{split}
s_E&=\left[\mathbf{P^I}; \mathbf{r}^{I}; \mathbf{d}^{I}; \mathbf{P^R}; \mathbf{r}^{R}; \mathbf{d}^{R}; \mathbf{\iota}; \mathbf{e}; \right],~ \\
s&=\left[\mathbf{r}^{I}; \mathbf{d}^{I}; \mathbf{r}^{R}; \mathbf{d}^{R}; \mathbf{e}\right].~
\end{split}
\end{equation}
where $\mathbf{P^I}\in [0, 1]^n$ ($\mathbf{P^R}\in [0, 1]^n$) is a vector where each element indicates the probability of a user changing e-state to \textit{Infected} (\textit{Recovered}), $\mathbf{r}^I\in \mathbb{Z}_{2}^{n}$ ($\mathbf{r^R}\in \mathbb{Z}_{2}^{n}$) is a vector indicating for each of the $n = |U|$ users whether they are in e-state \textit{Infected} (\textit{Recovered}) or not, $\mathbf{d}^I\in \mathbb{R}^{n}$ ($\mathbf{d}^R\in \mathbb{R}^{n}$) is a vector indicating the number of times each user had propagated fake news (true news) since the beginning of fake news mitigation, $\mathbf{\iota}\in \mathbb{R}^{n}$ is a vector indicating the intensity of each user, and $\mathbf{e}\in \mathbb{R}^{n}$ is a vector indicating the number of followers of each user. The action, $a$, to be chosen at each stage is which user $i \in U$ is to be selected as the debunker. 

To learn an optimal fake news mitigation policy under this reinforcement learning framework, we face a key challenge that the reward for selecting debunkers is not immediately available. Due to the continual and interleaving nature of information propagation on social networks, the net effect of selecting a debunker to post true news on the network is not directly and immediately observable; instead only the {\em network effect} can be seen at the end of a campaign. Thus, we define the episodic reward of an episode $\tau$ as:
\begin{equation}
\label{eq6}
V\left(\tau\right)=-log(\frac{||\mathbf{r}^I_{\tau}||_1}{n}).
\end{equation}
where $\mathbf{r}^I_{\tau}\in \mathbb{Z}_{2}^{n}$ is a vector indicating whether each of the $n$ users is in e-state \textit{Infected} after the fake news mitigation finishes. There are two possible methods to decide whether a user is \textit{Infected}: First, if a user shares a piece of fake news, the user is \textit{Infected}. Second, a user is \textit{Infected} based on the probability provided by Eq \ref{eq:infected}. In this paper, we use the second method. The episodic reward is measured at time $t_f$ after the multi-stage fake news mitigation ends. 

\section{Methodology}\label{sec:method}
Under the reinforcement learning framework, multi-stage fake news mitigation campaigns have episodic rewards. One approach to addressing episodic rewards is self-imitation learning~\citep{oh2018self,guo2018generative}. These methods imitate the past good behaviours of the learner itself, using the signal provided by the episodic reward, and can learn strong policies. 

In this section, we first introduce the self-imitation learning framework and then describe our approach, NAGASIL, namely {\underline N}egative sampling and state {\underline A}ugmented {\underline Generative} {\underline Adversarial} {\underline S}elf-{\underline I}mitation {\underline L}earning.

\subsection{Self Imitation Learning}
Originating from applications such as self-driving cars and drone manipulation, the imitation learning framework learns a policy to produce episodes similar to those of a human demonstrator, rather than learning from feedback on agent actions at intermediate steps~\citep{BojarskiTDFFGJM16,ross2013learning,HesterVPLSPSDOA17,VecerikHSWPPHRL17}. The same framework is used to derive Generative Adversarial Imitation Learning (GAIL), which is motivated by minimizing the divergence between the agent’s rollouts and expert demonstrations \citep{ho2016generative}. GAIL has been extended to GASIL (Generative Adversarial Self-Imitation Learning)~\citep{guo2018generative} by replacing expert data with the past good experiences of the learner itself (i.e., episodes with large episodic reward), which is a form of self-imitation learning. Given a policy $\pi$, occupancy measure $\rho_\pi$ is the distribution of state-action pairs that an agent encounters when navigating the environment under this policy. GAIL finds a policy $\pi$ whose occupancy measure $\rho_\pi$ minimizes the Jensen-Shannon divergence to the distribution of state-action pairs in past good experiences, i.e., it minimizes $D_{JS}(\rho_\pi,\rho_{\pi_E})$ where $\pi_E$ is the mixture policy represented by past good experiences. The causal entropy $H(\pi) \triangleq \mathbb{E}_\pi [-\log \pi(a|s)]$ is included as a policy regularizer to guard against collapse to a deterministic policy. To improve exploration, an ensemble of self-imitating agents is explicitly encouraged to visit different, non-overlapping regions of the state-action space, i.e., to simultaneously learn multiple diverse policies that explore different regions of the task~\citep{gangwani2018learning}. 

\begin{figure}[tb]
 \includegraphics[width=\linewidth]{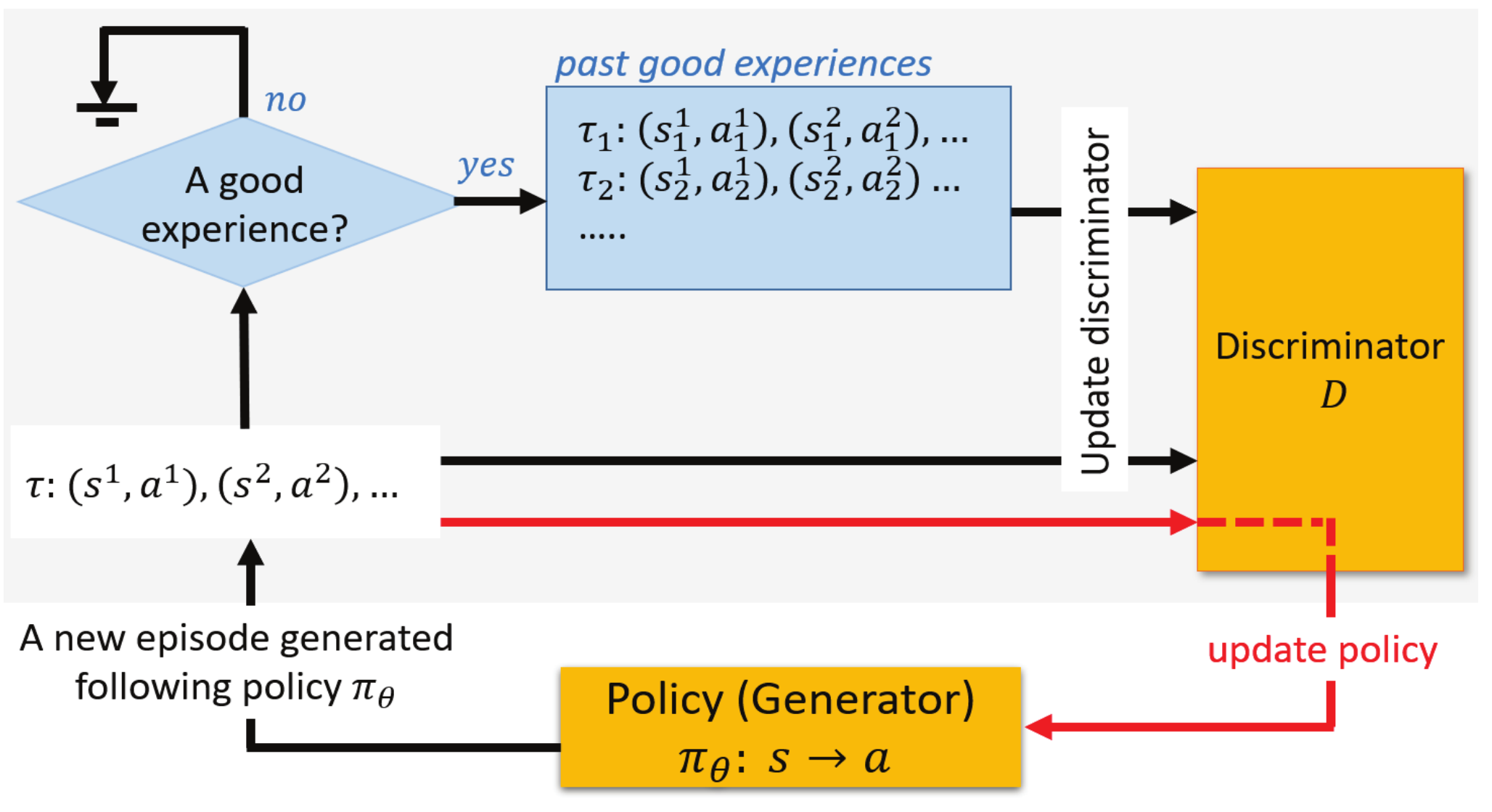} 
 \caption{Generative Adversarial Self-Imitation Learning.}\label{fig:awesome_image2}
\end{figure}

Fig.~\ref{fig:awesome_image2} illustrates GASIL. Episodes are generated following policy $\pi_{\theta}$ parameterised by $\theta$. In fake news mitigation, each episode is a sequence of $(s, a)$ pairs where $a$ is the debunker selected at state $s$. The episodes with the highest reward are considered as past good experiences. Higher reward means that the number of users in e-state \textit{Infected} after the fake news mitigation finishes, is smaller. Other experiences will be discarded during the learning process. Discriminator $D_\phi(s,a): S\times A\rightarrow[0, 1]$ is parameterized by $\phi$. The policy is trained to select action $a$ for state $s$ in a similar way to the mixture policy $\pi_E$ represented by past good experiences. The discriminator is trained to evaluate the discrepancy between the distribution of $(s, a)$ generated following policy $\pi_\theta$ and the distribution of $(s, a)$ in past good experiences. The policy and discriminator are improved alternately. To summarise, the aim is to find a policy $\pi_\theta$ whose occupancy measure $\rho_{\pi_\theta}$ minimizes the Jensen-Shannon divergence to the distribution of state-action pairs in past good experiences, i.e., to minimize $D_{JS}(\rho_\pi,\rho_{\pi_E})$~\citep{ho2016generative,guo2018generative,gangwani2018learning}. 

\subsection{NAGASIL}
We next describe our algorithm, NAGASIL, for learning a policy for multi-stage fake news mitigation on social networks in a sample-efficient manner.

\textbf{Negative Samples.} Fake news mitigation is cost-intensive, meaning that we aim to learn a policy with a limited number of episodes given some budget constraints. To this end, the strategy of negative sampling is explored. 
With past bad experiences only, we train a machine learning model $\mathcal{M}$ to predict the probability of actions being selected in state $s$. The output is a vector denoted as $\mathcal{M}(A|s)$ where each element corresponds to one action. Let $\pi_{\theta}(A|s)$ be the vector indicating the probability of each action following policy $\pi_{\theta}$. If $\mathcal{M}(A|s)$ indicates an action with a probability higher than that of the same action in $\pi_{\theta}(A|s)$, it implies we should avoid that action. We define the negative samples regularizer: 
\begin{equation}\label{eq:negative_regularizer}
\mathcal{N}(\pi_\theta)\triangleq \|\pi_{\theta}(A|s)-\mathcal{F}(\pi_{\theta}(A|s)-\mathcal{M}(A|s))\|_2^2.    
\end{equation}
where $\mathcal{F}(V)$ returns 0 for each negative element and returns the value in $\pi_\theta(A|s)$ for each positive element. That is, the regularizer punishes the negative elements but ignores the positive elements. The effectiveness of negative samples is analyzed in Proposition~1, where it is shown that the policy learned from negative samples on top of good experiences is guaranteed to be at least as good as the policy learned with only good experiences.

\begin{proposition}~\\
Provided that low Q-value state-action pairs appear more often in past bad experiences, $\mathbb{E}[\pi_{\theta}(a|s, s')Q(s, s', a)]\geq \mathbb{E}[\pi_{\theta_1}(a|s, s')Q(s, s', a)]$ where $\pi_{\theta}$ is learnt with the negative samples while $\pi_{\theta_1}$ is learnt without the negative samples. Proof can be found in Appendix \ref{proof:pro1}.
\end{proposition}

\textbf{Augmented State.} 
As the information propagation on social networks is interleaved, it is impossible to observe all the features needed for action selection. To address this issue, we propose to augment the state $s$, derived from the observable information, with all previous state-action pairs from the same episode. The rationale is to include as much historical information in the input as possible so that the agent can infer missing features in the observable state based on historical data. The augmented state $s'$ is defined as follows:
\begin{equation}\label{eq_aug_s}
s'_{i+1}=\frac{1}{i}\sum_{m=1}^{i} \psi^{i-m}\left[s_m; a_m\right].
\end{equation}
Here $\psi \in [0, 1]$ is the discount rate. Instead of finding policy $\pi_{\theta}(a|s)$, we aim to find policy $\pi_{\theta}(a|s, s')$.

Proposition~2 provides a theoretical justification of the augmented state, showing that the policy trained with the augmented state will be at least as strong as the policy trained otherwise.

\begin{proposition}~\\
$\mathbb{E}[\pi_{\theta_1}(a|s, s')Q(s, s', a)]\geq\mathbb{E}[\pi_{\theta_2}(a|s)Q(s, a)]$ where $\pi_{\theta_1}$ is trained with the augmented state while $\pi_{\theta_2}$ is trained without the augmented state. Both policies are learnt via Eq. \ref{eq:L} without negative samples. $Q(s, a)$ is the action-value function. Proof can be found in the Appendix \ref{proof:pro1}.
\end{proposition}

We integrate the negative samples and augmented state with GASIL, such that the objective of the discriminator and generator is defined as: 
\begin{equation}\label{eq:L}
\begin{split}
    \arg\min_{\theta}\arg\max_{\phi} \mathcal{L}= \mathbb{E}_{\pi_{\theta}}[\log D_\phi(s, s', a)]\\ + \mathbb{E}_{\pi_E}[\log (1-D_\phi(s, s', a))]-\lambda \mathcal{H}(\pi_{\theta}) + \lambda_1 \mathcal{N}(\pi_\theta).
\end{split}
\end{equation}
where $\mathcal{H}(\pi) = \mathbb{E}[-\log\pi(a|s)]$ is causal entropy~\citep{ho2016generative}. The optimized $\theta$ and $\phi$ minimize $D_{JS}(\rho_\pi,\rho_{\pi_E})$ which is:
\begin{equation}
\begin{split}
D_{JS}(\rho_{\pi_\theta},\rho_{\pi_E}) =\max_{\phi}\mathbb{E}_{(s, s', a)\sim\rho_{\pi_\theta}}[\log D_\phi(s, s', a)]\\ + \mathbb{E}_{(s, s', a)\sim\rho_{\pi_E}}[\log (1-D_\phi(s, s', a))].
\end{split}
\end{equation}

\begin{algorithm}[tb]
\caption{Selecting Debunkers via NAGASIL} 
\label{alg:gasil}
\begin{algorithmic}[1] 
\item Initialize discriminator $D$ with random parameters $\phi$, policy $\pi$ with random parameters $\theta$
\item Initialize past good experience memory $M_E$, past bad experience memory $M_B$
\For{iterations}
    \State Generate episode $\tau_{\pi_\theta}$ following policy $\pi_{\theta}$ \label{line:4}
    \State Update past good and bad experience memory $M_E$ and $M_B$ using $\tau_{\pi_\theta}$ \label{line:5}
    \State Sample minibatch $\tau_{E}$ from $M_E$
    \State Sample minibatch $\tau_{B}$ from $M_B$
    \State Update machine learning model $\mathcal{M}$ with $\tau_{B}$
    \State Update $D$ by ascending stochastic gradient $\nabla_{\phi} \mathcal{L}$ \label{line:D} with $\tau_{E}$ and $\tau_{\pi_\theta}$
    \State Update $\pi$ by ascending stochastic gradient $\nabla_{\theta} \mathcal{L}$\label{line:pi} with $\tau_{\pi_\theta}$
\EndFor 
\end{algorithmic} 
\end{algorithm}

\textbf{Selecting Debunkers via NAGASIL.} Our algorithm for fake news mitigation is presented in Alg. \ref{alg:gasil}. Following policy $\pi_\theta$, the generated episode $\tau_{\pi_\theta}$ is evaluated; if the episodic reward is one of the highest, $\tau_{\pi_\theta}$ is inserted into $M_E$; if it is one of the lowest, $\tau_{\pi_\theta}$ is inserted into $M_B$ (line \ref{line:4}-\ref{line:5}). 
The parameters $\phi$ of discriminator $D$ are updated (line \ref{line:D}) via stochastic gradient ascent:  
\begin{equation}
\begin{split}
\nabla_{\phi} \mathcal{L} = \mathbb{E}_{\tau_{\pi_\theta}\sim\pi_{\theta}}\left[\nabla_{\phi} \log D_{\phi}(s, s', a)\right] \\+\mathbb{E}_{\tau_{E}\sim\pi_E}\left[\nabla_{\phi} \log \left(1-D_{\phi}(s, s', a)\right)\right].
\end{split}
\end{equation}
The parameters $\theta$ of policy $\pi$ are updated (line \ref{line:pi}) via stochastic gradient ascent:
\begin{equation}
\begin{split}
\nabla_{\theta} \mathcal{L}=\mathbb{E}_{\tau_{\pi_\theta}\sim\pi_{\theta}}[\nabla_{\theta}\log D_\phi(s, s', a)] \\-\lambda \nabla_{\theta}\mathcal{H}(\pi_{\theta}) + \lambda_1 \nabla_{\theta}\mathcal{N}(\pi_\theta).
\end{split}
\end{equation}
The algorithm continuously updates the policy, which takes the observed environment state as input, and outputs a selected user as a debunker for fake news mitigation. 

\section{Experiments}\label{sec:experiments}
We evaluated NAGASIL on both real-world rumour datasets and social networks,  as well as large synthetic networks, and benchmarked it against baselines from both the fake news mitigation and self-imitation learning literature. Experiments were conducted on a cluster where each node has 64 cores, 2.0Ghz CPUs and 256G RAM. All deep neural networks are implemented using Tensorflow~\citep{abadi2016tensorflow} (distributed with the Apache License 2.0), all social networks are implemented using NetworkX~\citep{hagberg2008exploring} (distributed with the 3-clause BSD License) and the epidemic model SEIR is implemented based on EoN~\citep{miller2020eon} (distributed with the MIT License).

\textbf{Data.}  
We use PHEME~\citep{zubiaga2016analysing}, a widely used dataset for modeling rumour propagation on social networks. PHEME includes the source and the propagation path of messages, i.e., who spreads the news first and how the news propagates on social networks. From PHEME, the probability distribution that users propagate received news is extracted and used in our experiments. As the public PHEME dataset does not include the underlying social network, we utilize the Facebook social network~\citep{leskovec2012learning} from SNAP~\citep{snapnets} and also generate several synthetic Twitter social networks according to Twitter network settings from the literature~\citep{myers2014information}.

\begin{figure*}
    \centering
\subfigure {
\includegraphics[width=1.8\columnwidth]{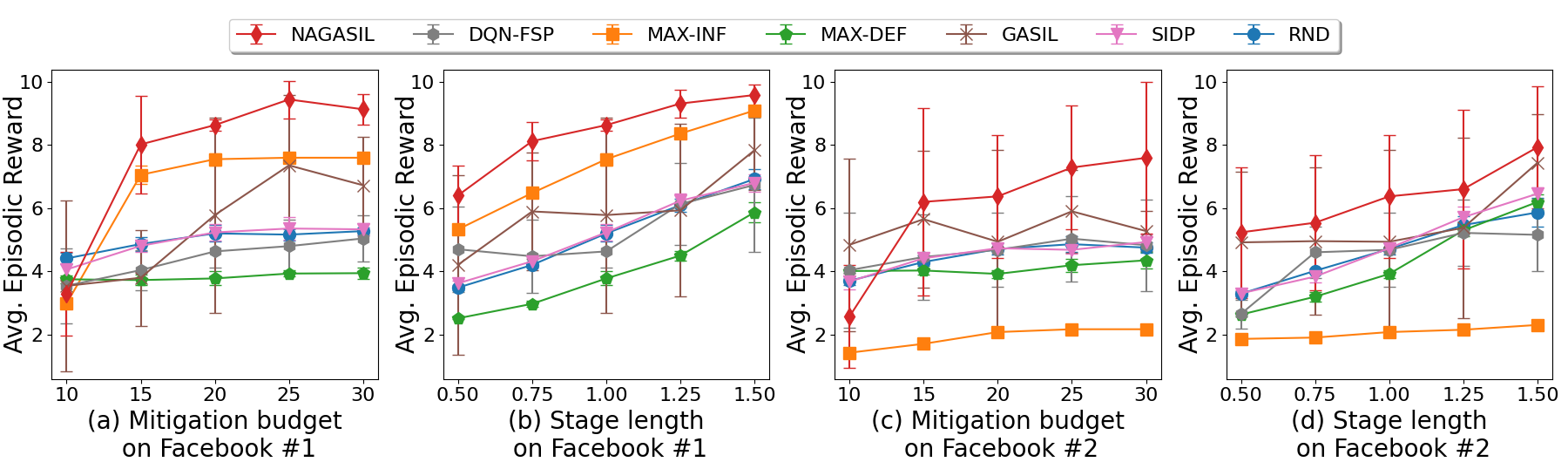} 
}     
    \caption{Performance for rumour mitigation on a real-world Facebook social network.}
    \label{fig:real-param}
\end{figure*}

\begin{figure*}
    \centering
\subfigure {
\includegraphics[width=1.8\columnwidth]{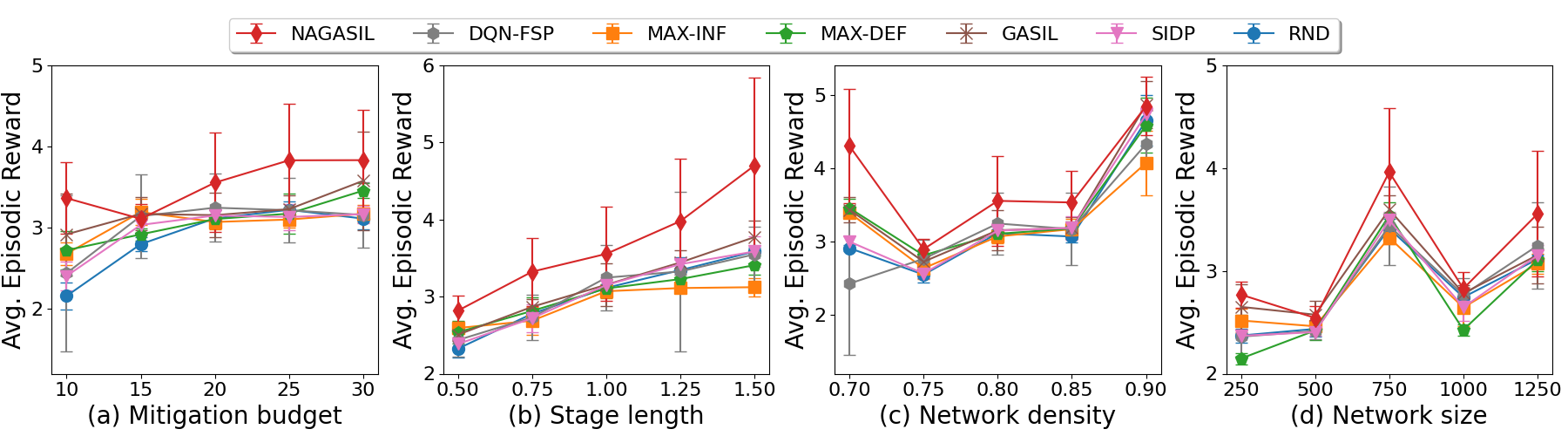} 
}     
    \caption{Performance for rumour mitigation on synthetic Twitter networks with various settings.}
    \label{fig:syn-param}
\end{figure*}

\textbf{Baselines and Evaluation Metrics.}\label{evaluationmetrics} We compare our proposed NAGASIL against baselines from both the fake news mitigation and reinforcement learning literature. Each method selects a debunker for every stage -- $w$ time steps -- until a given budget is used up. For each method, performance is evaluated based on the episodic reward of the episodes generated following the policy in the testing stage (i.e., the last 100 episodes). The episodic reward is defined by Eq. \ref{eq6}. Three baselines are state-of-the-art models from the fake news mitigation literature: 
\begin{itemize}
	\item \emph{Deep Q-Network with Future State Predictor} (DQN-FSP)~\citep{xiaofei2022} is a state-of-the-art RL-based method for fake news mitigation. When training DQN-FSP with single debunker selection, a reward will be provided following Eq. \ref{eq6}.
    \item \emph{Maximum Influence} (MAX-INF)~\citep{saxena2020k} is a widely used heuristic approach in the fake news mitigation literature. It selects the user with the highest number of followers (i.e., the most influential users) as the debunker for each stage.
	\item \emph{Maximum Defense} (MAX-DEF)~\citep{saxena2020k} is another heuristic approach for fake news mitigation. It selects the most active user propagating fake news as the debunker for each stage. 
\end{itemize}
Two more baselines can be formed by applying existing self-imitation learning models for fake news mitigation: 
\begin{itemize}
	\item \emph{Generative Adversarial Self-Imitation Learning} (GASIL) is a popular self-imitation learning method proposed in \citep{guo2018generative} (source code distributed with MIT License).
	\item \emph{Self-Imitating Diverse Policies} (SIDP) is a self-imitation learning method, improved to achieve better exploration of the environment~\citep{gangwani2018learning}.
\end{itemize}
As a sanity check, we include as a baseline the basic policy of randomly selecting one user as the debunker at each stage (RND). Note that we have experimented with other general RL-based methods, such as PPO and DQN, but they perform consistently worse than RND and thus do not pass our sanity check. A possible reason for this is that these methods are not designed for episodic reward settings. Results for these methods, therefore, are not reported. The details of experiment settings are presented in Appendix \ref{sec:syn_parm}.

\subsection{Performance w.r.t. Mitigation Settings}
Figure \ref{fig:real-param} and \ref{fig:syn-param} report the average episodic reward of five runs using different random seeds and their standard deviation for PHEME propagation on a real-world Facebook social network and (synthetic) Twitter networks with different settings. From the Facebook data, we randomly pick two ego networks, named Facebook \#1 and \#2. Figure \ref{fig:real-param}(a) and (c) show the performance with respect to the budget $B$, where the budget increases from $10$ to $30$, on Facebook networks. Given more budget, more mitigation effect is expected, since the fake news mitigation campaign will continue for more stages. The experiment results show that NAGASIL performs better on all budget settings except on extremely low budgets. This might be caused by the extremely limited number of actions, where the problem effectively reduces to one-off selections. Figure \ref{fig:real-param}(b) and (d) show the performance with respect to stage length $w$ on Facebook networks. The greater stage length allows the agent to observe more mitigation effects from previous actions; however, the greater stage length means it takes a longer time to select the next debunker. To investigate the impact caused by stage length, we increase the stage length from $0.5$ to $1.5$. We can see that NAGASIL has outstanding performance in all settings against all baselines. MAX-INF demonstrates much better performance in Facebook \#1 than in Facebook \#2. It suggests that the performance of MAX-INF is unstable, i.e., it is heavily influenced by the underlying social network structure.

Figure \ref{fig:syn-param}(a) shows the performance with respect to the budget $B$ on a Twitter network of 1250 nodes (see Appendix ~\ref{sec:syn_parm}). We also vary the budget $B$ from $10$ to $30$. The experiment results demonstrate that NAGASIL performs consistently better under most budget settings. Figure \ref{fig:syn-param}(b) shows the performance with respect to stage length $w$, varying from $0.5$ to $1.5$ on the Twitter network. We can see that NAGASIL has consistently outstanding performance in all settings against all baselines. 

\subsection{Performance w.r.t. Network Settings}\label{sec:envset}
Figure \ref{fig:syn-param}(c) shows the performance with respect to Twitter networks of different densities. In the network generation model \citep{bollobas2003directed}, the density was controlled by $\beta$ (by default, $\beta = 0.8$). In the experiments, $\beta$ is increased from $0.7$ to $0.9$ to simulate different levels of density. The sum of $\alpha$, $\beta$ and $\gamma$ must be 1. We keep $\gamma = 3\alpha$ to ensure that the social networks maintain the property that the out-bound degree is lower than the in-bound degree~\citep{myers2014information}. We can observe the significant advantage of NAGASIL against all baselines at most density levels. Figure \ref{fig:syn-param}(d) shows the performance with respect to the size of the Twitter networks where the number of users increases from $250$ to $1250$. The density of Twitter networks is the same by setting $\beta = 0.8$. Clearly, NAGASIL outperforms all baselines consistently. 

\begin{figure}[tb]
    \centering
\subfigure[Ablation study] {
 \label{fig:ablation}     
\includegraphics[width=0.6\linewidth]{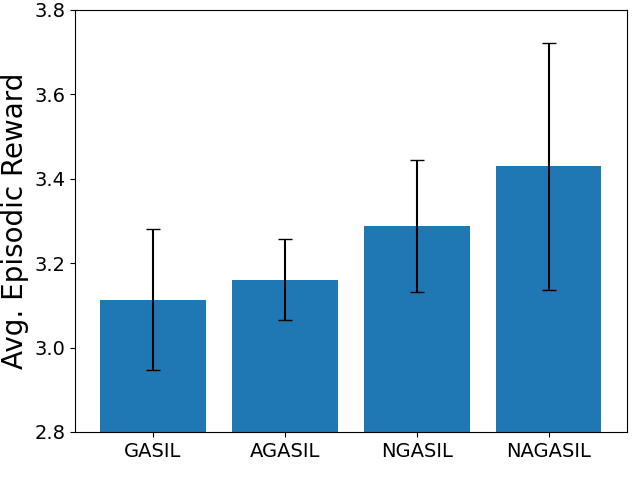} 
}
\subfigure[Larger networks test] {
 \label{fig:large_network}     
\includegraphics[width=0.64\linewidth]{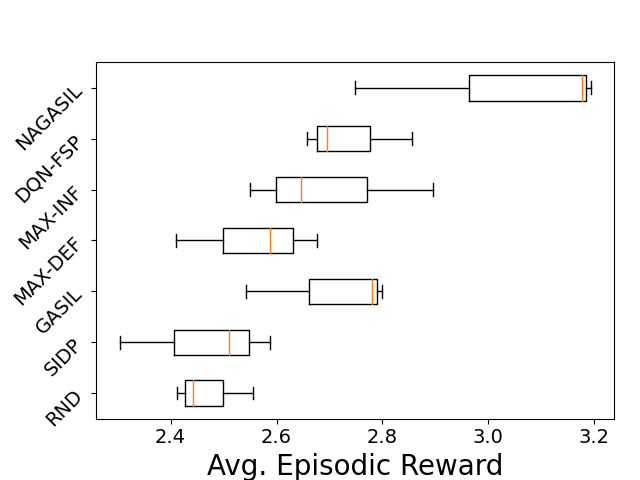}
}
    \caption{NAGASIL - ablation study and performance on larger networks.}
    \label{fig:discussion}
\end{figure}

\subsection{Ablation Study}\label{sec:ablation}
To verify the effectiveness of negative samples and augmented state, we compare the performance of NAGASIL against three ablated versions: NGASIL, AGASIL, and GASIL. NGASIL is a variant of NAGASIL where the augmented state is removed (i.e., $s'$ is removed in Eq. \ref{eq:L} when learning the policy). AGASIL is a variant of NAGASIL where the negative samples are removed (i.e., $\lambda_1$ is set to 0 in Eq. \ref{eq:L} when learning the policy). The default experiment settings on Twitter networks (see Appendix \ref{sec:syn_parm}) are applied except the evaluation metric is calculated over all 1000 episodes to demonstrate the performance of different baselines during the whole training process. 

The results of the ablation study are reported in Figure \ref{fig:ablation}, where the mean and standard deviation of 5 runs are plotted for each method. Firstly, note that NGASIL has significantly better performance than GASIL, indicating that negative samples can clearly boost performance for fake news mitigation. Secondly, AGASIL has better performance compared to GASIL. Even though $s'$ provides additional information, the augmented state increases the state space. A larger state space necessitates more training data to fully realise the benefits brought by additional information. Thirdly, NAGASIL has a significant performance advantage compared to NGASIL, AGASIL and GASIL. The performance gain is attributed to the combination of negative samples and the augmented state; that is, the negative samples provide extra training data to help improve the effectiveness of the augmented state. 

\subsection{Performance on Larger Twitter Networks}
To simulate real-world deployment of fake news mitigation campaigns,  
we further evaluate NAGASIL on a larger Twitter network with 2500 users. The average performance of 3 runs of NAGASIL (default settings) and baselines is reported as a box plot in Figure~\ref{fig:large_network}. The experiment results clearly show that NAGASIL outperforms all baselines on the larger Twitter network. Note that following real-world fake news mitigation reported in the literature~\citep{farajtabar2017fake}, our setting simulates that mitigation policy is applied to a group of users who have interactions with the news to be mitigated, which often are in the order of thousands rather than millions. 

\section{Limitations}
Our NAGASIL framework provides a data-driven platform for media researchers to conduct studies and discover factors affecting the interaction between misinformation and truthful information propagation on social networks. However, similar to existing studies, e.g., \citep{farajtabar2017fake,xiaofei2022}, our proposed fake news mitigation policy assumes that the truth value -- true or fake -- of social media news posts is established and fed to the mitigation process. Therefore, the mitigation model must be applied in conjunction with a fake news detection model. 
 
\section{Conclusion}
This study identified and addressed the issue of \textit{episodic reward}, an essential but overlooked issue for learning fake news mitigation policies via reinforcement learning. Specifically, our solution proposed the \textit{negative samples} to learn the policy for debunker selection by imitating past good experiences and avoiding past bad experiences. We further proposed the \textit{augmented state} to enrich the state of reinforcement learning with features extracted from past state-action pairs in the current episode. The superiority of the proposed solution has been verified against various baselines in different settings. In a broader context, this study improves fake news mitigation, for the first time to our best knowledge, by giving due consideration to the complexity of information propagation on social networks. While NAGASIL is compatible with different information propagation models, a single propagation model cannot fully characterise the complex information propagation on social networks. In future work, we will further investigate NAGASIL with alternative information propagation models to further evaluate its robustness.

\section*{Acknowledgements}
This research is supported in part by the ARC Discovery Projects DP200101441, DP210100743 and ARC Linkage Project LP180100750.
%
%
\bibliographystyle{abbrvnat}
\bibliography{bibfile}

\newpage
\appendix
\section{Proof of Propositions}
\label{proof:pro1}
\subsection{Proposition 1.}
\textit{$\mathbb{E}[\pi_{\theta_1}(a|s,s')Q(s,s',a)]\geq\mathbb{E}[\pi_{\theta_2}(a|s)Q(s,a)]$ where $\pi_{\theta_1}$ is trained with the augmented state while $\pi_{\theta_2}$ is trained without the augmented state. Both policies are learnt via Eq. \ref{eq:L} without negative samples. $Q(s,a)$ is the action-value function.}

\emph{proof:} $\rho_{E}(s, a)$ can be calculated by summing the joint probability distribution over all values of $s'$, that is, $\rho_{E}(s, a) = \sum_{s'}\rho_{E}(s, s', a)$. By definition, $\rho_{E}(s, a)=P(s)\pi_E(a|s)$ and $\rho_{E}(s, s',a )=P(s)\sum_{s'}\pi_E(a|s, s')$ where $P(s)$ is probability of $s$ in past good experiences. Since $\pi_E(a|s)= \sum_{s'}\pi_E(a|s, s')$, we have $\pi_E(a|s)\geq\pi_E(a|s, s')$ which indicates $a$ is equally or more preferable if considering $s$ than that if considering $(s, s')$. As a result, $\rho_E(s, a)\geq\rho_E(s, s', a)$ where $\rho_E(s, a)$ includes the probability of $(s, s', a)$ inside, and outside, past good experiences.  According to~\citep{ho2016generative}, $\rho_{\pi_{\theta_1}}=\rho_{E}(s, s', a)$ and $\rho_{\pi_{\theta_2}}=\rho_{E}(s, a)$ by learning policy via Eq. \ref{eq:L} without $\mathcal{N}(\pi_\theta)$. The episodes generated following policy $\pi_{\theta_1}$ are equally or more likely to have high rewards compared with those generated following $\pi_{\theta_2}$. That is, $\mathbb{E}[\pi_{\theta_1}(a|s, s')Q(s, s', a)]\geq\mathbb{E}[\pi_{\theta_2}(a|s)Q(s,a )]$. \qed 

\label{proof:pro2}
\subsection{Proposition 2.}
\textit{Provided that low Q-value state-action pairs appear more often in past bad experiences, $\mathbb{E}[\pi_{\theta}(a|s, s')Q(s, s', a)]\geq \mathbb{E}[\pi_{\theta_1}(a|s, s')Q(s, s', a)]$ 
where $\pi_{\theta}$ is learnt with the negative samples while $\pi_{\theta_1}$ is learnt without the negative samples.} 
 
\emph{proof:} $\pi_\theta(a|s, s')$ and $\pi_{\theta_1}(a|s, s')$ are updated in the same way when $\rho_{\hat{E}}(s, s', a)$ changes. Let $\rho_{\hat{E}}$ be the occupancy measure of past bad experiences. If $\rho_{\hat{E}}(s, s', a)=0$, $\pi_{\theta}(a|s, s')$ is same as $\pi_{\theta_1}(a|s, s')$. If $\rho_{\hat{E}}(s, s', a)>0$, when it becomes relatively greater (or less) than $\rho_E(s, s', a)$, the Q-value of corresponding state-action pair $Q(s, s', a)$ will be less (or greater). Accordingly, $\pi_{\theta}(a|s, s')$ will be reduced (or increased); but, $\pi_{\theta_1}(a|s, s')$ will not change since it is unaware relative change between $\rho_{\hat{E}}(s, s', a)$ and $\rho_E(s, s', a)$. Thus, $\mathbb{E}[\pi_{\theta}(a|s, s')Q(s, s', a)]\geq \mathbb{E}[\pi_{\theta_1}(a|s, s')Q(s, s', a)] $. \qed \\

The assumption that low Q-value state-action pairs appear more often in bad experiences is likely to hold much of the time since past bad experiences only include worst-performing episodes.

\section{Experiment Settings}\label{sec:syn_parm}
In this study, unless stated otherwise, all methods run for 1000 episodes, where the last 100 episodes are in the testing stage. By default, the experiments on real-world networks will randomly pick two ego networks, named Facebook \#1 and \#2, (with radius = 2, anonymized networks released with code) from the Facebook dataset~\citep{leskovec2012learning}. The experiments on synthetic Twitter social networks will have a directed scale-free network with $1250$ users generated using the method in~\citep{bollobas2003directed} (no self-loop and multiple edges). The parameters of network generation model are $\alpha = 0.05, \beta = 0.8$ and $\gamma = 0.15$. Each user $i$ has a cost $(e_i/\max_{j\in U}(e_j))*9+1$ which is in $[1, 10]$. Users with a higher number of followers will have higher costs. 20 users are randomly picked up as fake news spreaders who are in e-state \textit{Infected} with intensity $\xi_i \sim \mathcal{U}[0.5, 1.5]$ to spread fake news at time 0 and the intensity decays along with time following Eq. \ref{eq:intensity}; other users are in e-state \textit{Susceptible}. At time 5, fake news mitigation starts. For every time period $w=1$ (i.e., a stage), one user from $U$ is selected as a new debunker until budget $B=20$ is used up. If user $i$ is selected as a debunker, their e-state is changed to \textit{Recovered} with intensity $\xi_i \sim \mathcal{U}[0.5, 1.5]$ to spread true news immediately. After the last stage, the news spreading continues for a time period of 5 and then the episodic reward is gauged. For self-imitation learning baselines, the number of past good experiences is set to $20$, for NAGASIL, we additionally set the number of past bad experiences to $10\%$ of total past experiences.

When a user $i$ receives fake news (true news), the probability of changing from another e-state to \textit{Infected} (\textit{Recovered}) is updated according to Eq. \ref{eq:infected} (Eq. \ref{eq:recovered}) where the logistic function's midpoint $x_i$ in Eq. \ref{equ:lvalue} is defined as $(e_i/\max_{j\in U}(e_j))*2+1$, i.e., a value in $[1, 3]$. If user $i$ has more followers, $x_i$ is higher and the user is less likely to change e-state. Once user $i$ is successfully changed to e-state \textit{Infected} (\textit{Recovered}), the intensity of the user to spread fake news (true news) is initialized to be $\xi_i \sim \mathcal{U}[0.5, 1.5]$ and decays along with time following Eq. \ref{eq:intensity}. 
\end{document}